\documentclass[11pt]{article}
\usepackage{moriond,epsfig,amsmath,wrapfig}

\begin{document}

\pagestyle{empty}

\begin{flushleft}
\Large
{SAGA-HE-160-00
\hfill May 6, 2000}  \\
\end{flushleft}
 
\vspace{3.0cm}
 
\begin{center}
 
\huge{{\bf Parametrization of}} \\
\vspace{0.2cm}

\huge{{\bf polarized parton distributions}} \\

\vspace{1.5cm}
 
\LARGE
{\ S. Kumano $^*$} \\
 
\vspace{0.3cm}
  
\LARGE
{Department of Physics}         \\
 
\LARGE
{Saga University}      \\
 
\LARGE
{Saga 840-8502, Japan} \\

\vspace{2.0cm}
 
\LARGE
{Talk given at the XXXVth Rencontres de Moriond} \\

\vspace{0.1cm}

{on QCD and High Energy Hadronic Interactions} \\

\vspace{0.1cm}

{Les Arcs, France, March 18 -- 25, 2000} \\

\vspace{0.05cm}

{(talk on March 24, 2000) }  \\
 
\end{center}
 
\vspace{0.7cm}

\vfill
 
\noindent
{\rule{6.0cm}{0.1mm}} \\
 
\vspace{-0.3cm}
\normalsize
\noindent
{* Email: kumanos@cc.saga-u.ac.jp.
          Information on his research is available at} \\

\vspace{-0.44cm}
\noindent
{\ \ \ http://www-hs.phys.saga-u.ac.jp.}  \\

\vspace{+0.1cm}
\hfill
{\large to be published in proceedings}

\vfill\eject

\normalsize
%%%%%%%%%%%%%%%%%%%%%%%%%%%%%%%%%%%%%%%%%%%%%%%%%%%%%%%%%%%%%%%%%%%%%%%%%%%%%%%
%%%%%%%%%%%%%%%%%%%%%%%%%%%%%%%%%%%%%%%%%%%%%%%%%%%%%%%%%%%%%%%%%%%%%%%%%%%%%%%

%\begin{document}

$\ \ \ $ 

\vspace{-0.7cm}

\vspace*{4cm}
\title{PARAMETRIZATION OF POLARIZED PARTON DISTRIBUTIONS}

\author{S. KUMANO}

\address{Department of Physics, Saga University \\ 
         Saga 840-8502, Japan}

\maketitle\abstracts{We report the polarized parton distributions
proposed by the Asymmetry Analysis Collaboration (AAC).
Using parametrized distributions at $Q^2$=1 GeV$^2$ and measured
$A_1$ data, we determine optimum polarized distributions in
the leading order (LO) and next-to-leading order (NLO).
We find that the polarized antiquark distribution
is not well determined particularly at small $x$.
It could lead to a rather small quark spin content
in comparison with usually-quoted values of 10$\sim$30\%.
In our analysis, it varies from 5\% to 28\% depending
on the small-$x$ extrapolation.
It is necessary to have small $x$ $(<10^{-3})$ data for precise
determination. In addition, the large-$x$ region should be also
studied for $\Delta \bar q$, which cannot be determined solely by $g_1$.
We propose three sets of distributions as the longitudinally
polarized parton distribution functions.}

%%%%%%%%%%%%%%%%%%%%%%%%%%%%%%%%%%%%%%%%%%%%%%%%%%%%%%%%%%%%%%%%%%%%%%%%%%%%%%%
\section{Introduction}\label{intro}

From the measurements of the proton's polarized
structure function $g_1$, we learned that the proton spin
cannot be understood in a simple quark model. It should be
interpreted mainly by a gluon contribution and/or effects of angular
momenta; however, the situation is not satisfactory for specifying
the major carrier of the proton spin.
Our study is intended to understand the current status of
polarized parton distributions by analyzing inclusive spin asymmetry
$A_1$ data prior to RHIC-Spin measurements.
Semi-inclusive data became available, but they are not accurate enough
to provide any significant constraint. 
Our group name AAC stands for Asymmetry Analysis
Collaboration, which consists of theorists and experimentalists.\cite{aac}
In this paper, we report our analysis of the asymmetry $A_1$ for getting
optimum polarized parton distributions. 

In Sec.\ref{para}, a parametrization form of the polarized parton
distributions is discussed. The results are shown in Sec.\ref{results}.
According to our analysis, the polarized antiquark distribution cannot be
well determined, and this issue is discussed in Sec.\ref{antiquark}.
The results in Sec.\ref{results} and \ref{antiquark} are quoted from Ref.1.
Finally, conclusions are given in Sec.\ref{concl}.

%%%%%%%%%%%%%%%%%%%%%%%%%%%%%%%%%%%%%%%%%%%%%%%%%%%%%%%%%%%%%%%%%%%%%%%%%%%%%%%
\section{Parametrization}\label{para}

First, an $x$-dependent functional form of parametrized
distributions is explained.
Because the positivity condition is imposed not only
in the leading order (LO) but also in the next-to-leading order (NLO),
it is convenient to have an unpolarized distribution multiplied by
an $x$-dependent function:
\begin{equation}
\Delta f_i(x, Q^2_0) = A_i \, x^{\alpha_i} \,
               (1 + \gamma_i \, x^{\lambda_i}) \, f_i(x, Q^2_0) ,
\label{eqn:PPDF2}
\end{equation}
at the initial $Q^2$ point ($Q_0^2$=1 GeV$^2$).
Since there is no accurate data to find the difference between
$\Delta \bar u$, $\Delta \bar d$, and $\Delta \bar s$, these distributions
are assumed to be the same. Of course, they are not expected to be
equal by considering the unpolarized situation.\cite{skpr}
With this assumption and the flavor number $N_f=3$,
we have four distributions ($\Delta u_v$, $\Delta d_v$, 
$\Delta \bar q$, $\Delta g$) to be determined by a $\chi^2$
analysis. Because there are four parameters for each distribution,
we have sixteen parameters in total. However, the first moments of
$\Delta u_v$ and $\Delta d_v$ are fixed ($\eta_{u_v}=0.926$ and
$\eta_{d_v}=-0.341$) by using semi-leptonic decay constants.
It means that the number of actual free parameters is fourteen.

The polarized distributions are evolved \cite{q2-evol}
to experimental $Q^2$ points of the spin asymmetry $A_1$,
which is expressed by the structure functions as
\begin{equation}
A_1(x,Q^2)\simeq \frac{2 \, x \, g_1 (x,Q^2) 
                         \, [ 1+R(x,Q^2)]}{F_2 (x,Q^2)}.
\label{eqn:a1}
\end{equation}
The longitudinal-transverse structure function ratio $R$ is taken
from the analysis of SLAC-1990, $F_2$ is calculated by 
the GRV98 distributions, and $g_1$ is calculated by our parametrized
distributions. 
Then, $\chi^2$ is evaluated in comparison with experimental data:
$\chi^2=\sum(A_1^{\rm data}(x,Q^2)-A_1^{\rm calc}(x,Q^2) )^2 /
 (\Delta A_1^{\rm data}(x,Q^2))^2$.
The optimum set of parameters is found by minimizing $\chi^2$
by the subroutine MINUIT.

%%%%%%%%%%%%%%%%%%%%%%%%%%%%%%%%%%%%%%%%%%%%%%%%%%%%%%%%%%%%%%%%%%%%%%%%%%%%%%%
\section{Results}\label{results}

\begin{figure}[h]
\parbox[b]{0.46\textwidth}{
   \begin{center}
      \epsfig{file=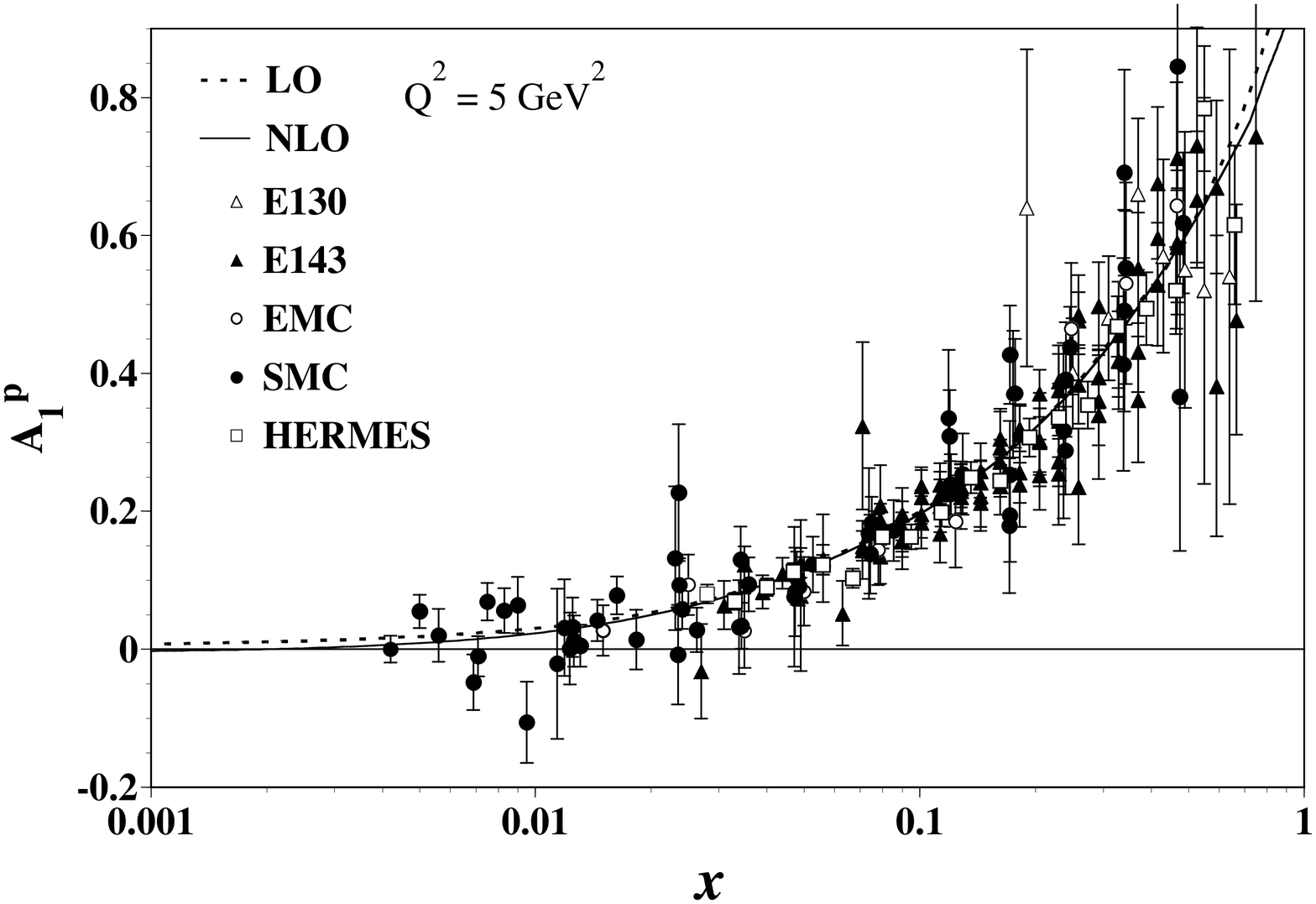,width=5.5cm}
   \end{center}
 \vspace{-0.6cm}
   \caption{\footnotesize Spin asymmetry $A_1$ for the proton.}
   \label{fig:a1p}
}\hfill
\parbox[b]{0.46\textwidth}{
   \begin{center}
      \epsfig{file=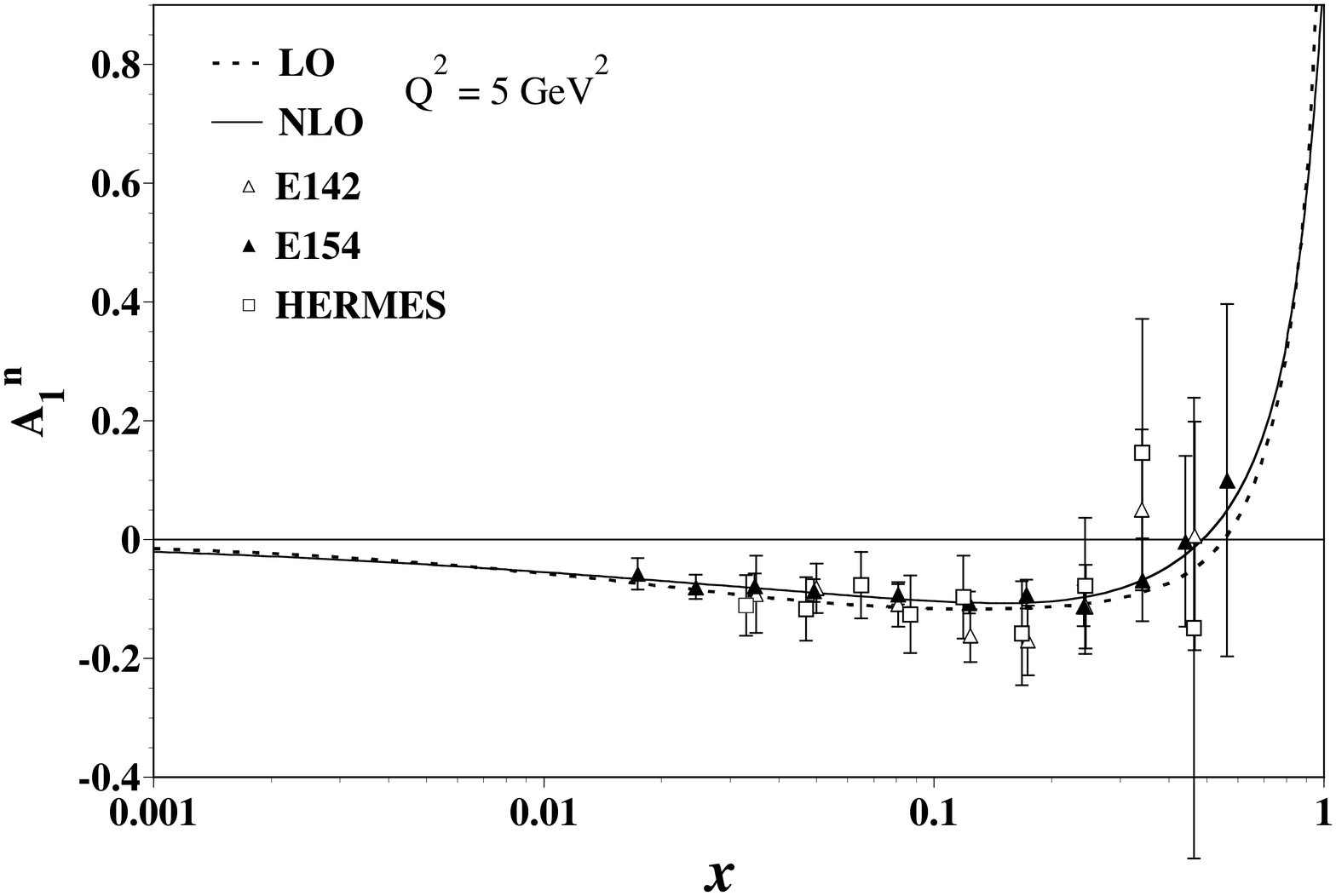,width=5.5cm}
   \end{center}
 \vspace{-0.6cm}
\caption{\footnotesize Spin asymmetry $A_1$ for the neutron.}
\label{fig:a1n}
}
\end{figure}

In our analysis, we use the asymmetry data set of so called ``large
tables''. On the other hand, other analyses use small data tables in which
$Q^2$ values are averaged at certain $x$ points.
Although the present data are not accurate
enough to provide $Q^2$-dependence information, we believe that
the raw data should be used as much as we can.
For 375 data points with $Q^2>1$ GeV$^2$, we obtain the minimum
$\chi^2$ values $\chi^2$/d.o.f.=322.6/360 and 300.4/360 for LO and NLO,
respectively. There is a significant $\chi^2$ reduction in NLO, so that
it is important to analyze the data by the NLO expressions.
It should be noted that all our NLO results are obtained
in the $\overline{\rm MS}$ scheme. There are two major sources for
the $\chi^2$ reduction, and they are from the HERMES proton
and E154 neutron data in Figs.\ref{fig:a1p} and \ref{fig:a1n}. 

\vspace{-0.5cm}
%%%%%%%%%%%%%%%%%%%%%%%%%%%%%%%% figure %%%%%%%%%%%%%%%%%%%%%%%%%%%%%%%%%%%%%%
\noindent
%%%%%%%%%%%%%%%%%%%%%%%%%%%%%%%% figure %%%%%%%%%%%%%%%%%%%%%%%%%%%%%%%%%%%%%%
\vspace{-0.2cm}

\noindent
In these figures, our optimum LO and NLO asymmetry
curves are shown at $Q^2$=5 GeV$^2$. Because the data are taken at
various $Q^2$ points depending on the $x$ region, it is not
straightforward to compare them with the fixed $Q^2$ curves.
However, the figures show that the agreement is satisfactory.
The errors of deuteron data are so large that its $A_1$ figure
is not shown here.

\begin{figure}[h]
\parbox[b]{0.46\textwidth}{
   \begin{center}
      \epsfig{file=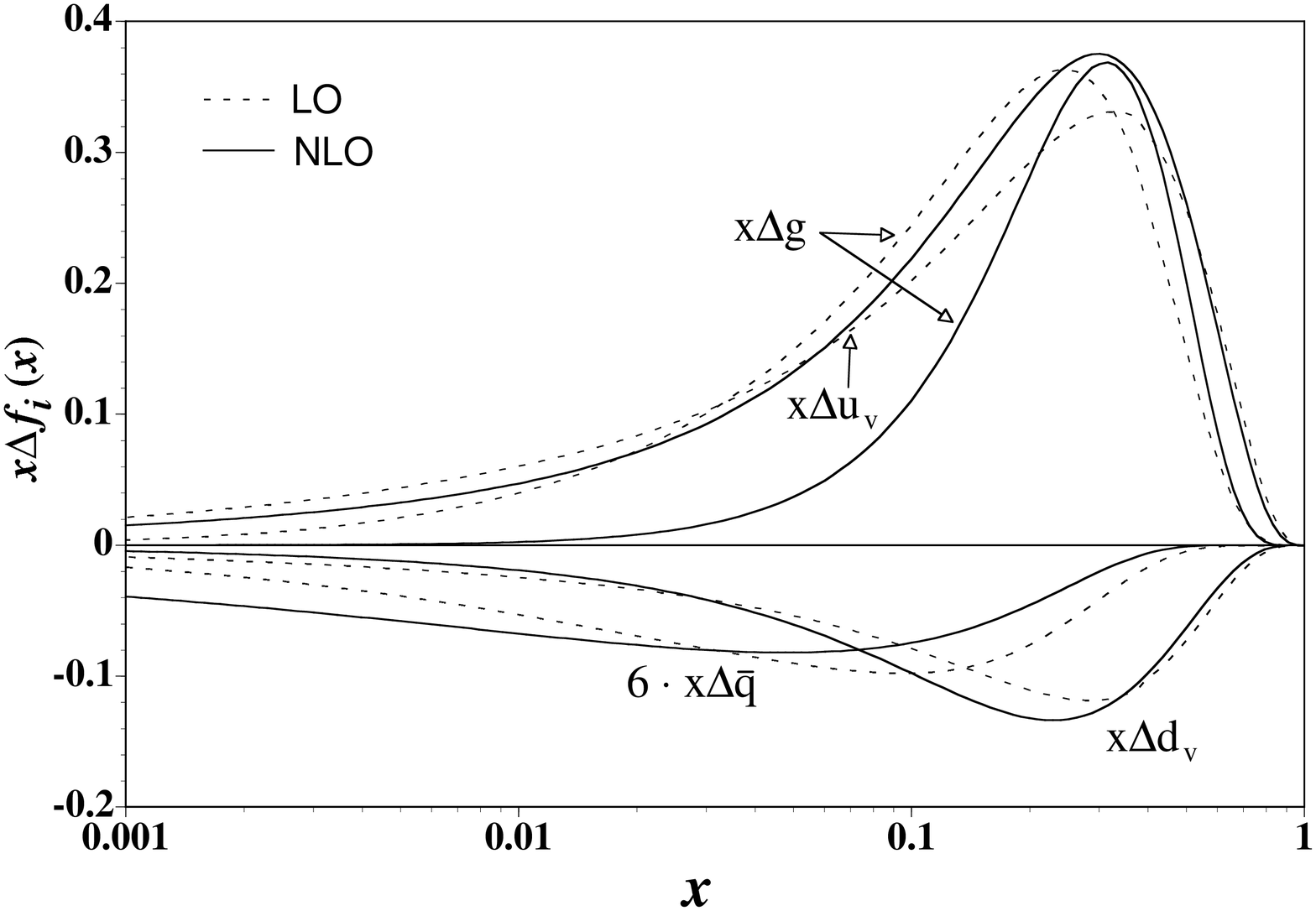,width=5.5cm}
   \end{center}
 \vspace{-0.6cm}
   \caption{\footnotesize Obtained parton distributions.}
   \label{fig:df}
}\hfill
\parbox[b]{0.46\textwidth}{
   \begin{center}
      \epsfig{file=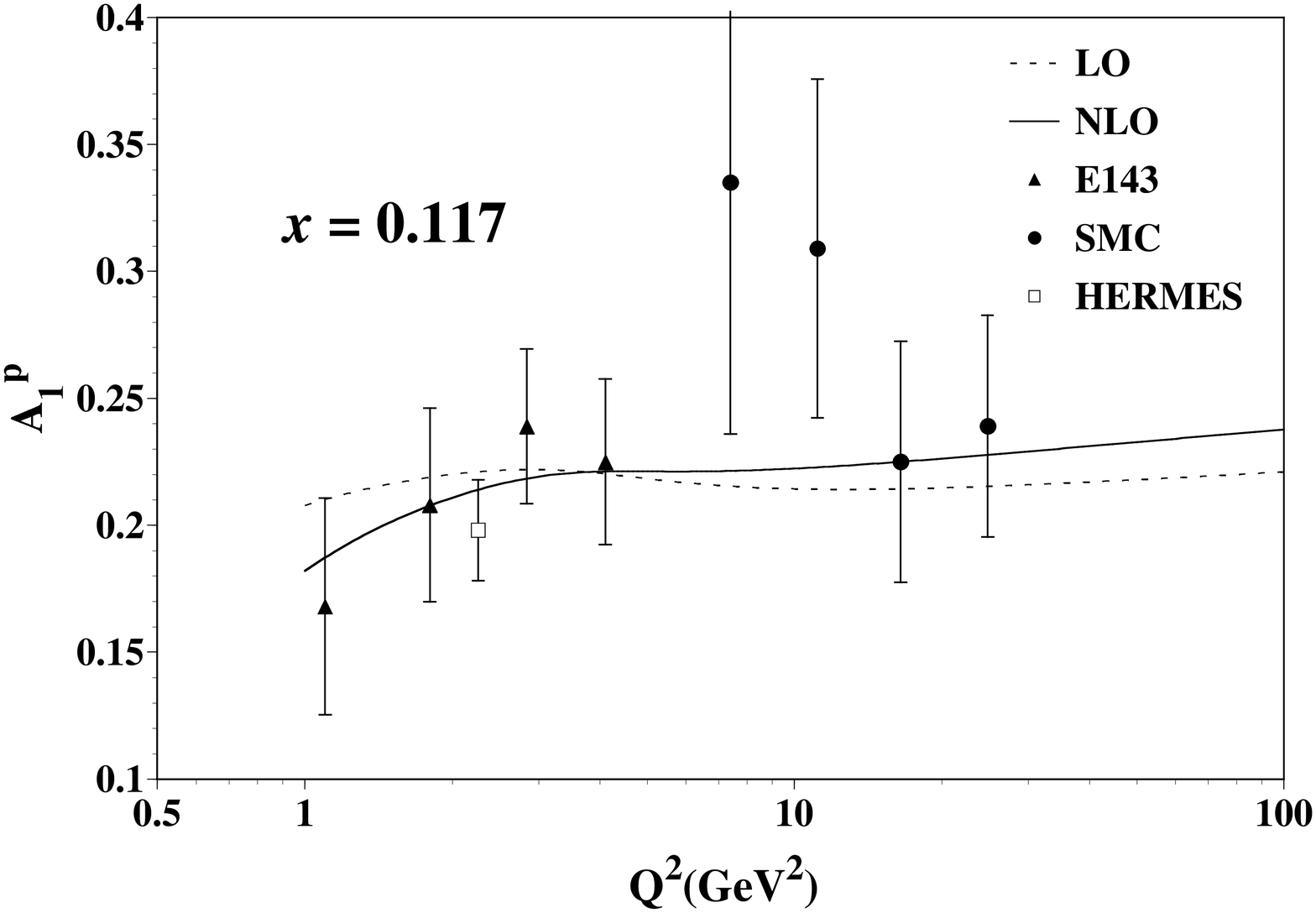,width=5.5cm}
   \end{center}
 \vspace{-0.6cm}
   \caption{\footnotesize $Q^2$ dependence of $A_1^p$.}
\label{fig:q2-a1}
}
\end{figure}

Obtained polarized parton distributions are shown in Fig.\ref{fig:df}.
Because the first moments of $\Delta u_v$ and $\Delta d_v$ are
fixed at positive and negative numbers, respectively, they are
mainly positive and negative distributions. It is rather difficult
to determine the gluon distribution, but a positive distribution
is favored in both LO and NLO. The antiquark distribution becomes
negative, and we discuss the issue of its determination in the next
section.

Using the obtained distributions, we show $Q^2$ dependence
of $A_1$ in comparison with data in Fig.\ref{fig:q2-a1}.
The distributions are provided at $Q^2$=1 GeV$^2$ and they are
evolved to larger $Q^2$ by the DGLAP equations. The curves show
that there could be strong $Q^2$ dependence in the small-$Q^2$ 
region ($Q^2<$2 GeV$^2$). Experimentalists sometimes assume 
that $A_1$ is $Q^2$ independent for evaluating $g_1$ from the
$A_1$ data. Although it does not matter due to the present
experimental accuracy, we should be careful about such an 
assumption for a precise analysis.

\vspace{-0.5cm}
%%%%%%%%%%%%%%%%%%%%%%%%%%%%%%%% figure %%%%%%%%%%%%%%%%%%%%%%%%%%%%%%%%%%%%%%
\noindent
%%%%%%%%%%%%%%%%%%%%%%%%%%%%%%%% figure %%%%%%%%%%%%%%%%%%%%%%%%%%%%%%%%%%%%%%
\vspace{-0.2cm}

Using these distributions, we get the quark spin content
\begin{equation}
\Delta\Sigma = 0.201 \ {\rm (LO)}, \ \ \  0.051 \ {\rm (NLO)}.
\end{equation}
The NLO value seems to be significantly smaller than
usually-quoted ones 10\%$\sim$30\%. 
The small spin content originates from the small-$x$ behavior of our
antiquark distribution, so that we discuss this issue in the next
section. 

%%%%%%%%%%%%%%%%%%%%%%%%%%%%%%%%%%%%%%%%%%%%%%%%%%%%%%%%%%%%%%%%%%%%%%%%%%%%%%%
\section{Polarized antiquark distribution}\label{antiquark}

%%%%%%%%%%%%%%%%%%%%%%%%%%%%%%%% figure %%%%%%%%%%%%%%%%%%%%%%%%%%%%%%%%%%%%%%
\begin{wrapfigure}{r}{0.46\textwidth}
   \vspace{-0.3cm}
   \begin{center}
      \epsfig{file=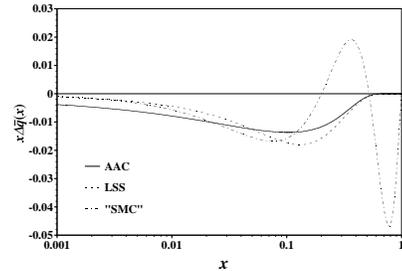,width=5.5cm}
   \end{center}
 \vspace{-0.7cm}
   \caption{\footnotesize Polarized antiquark distributions.}
   \label{fig:dqbar}
\end{wrapfigure}
%%%%%%%%%%%%%%%%%%%%%%%%%%%%%%%% figure %%%%%%%%%%%%%%%%%%%%%%%%%%%%%%%%%%%%%%
Because the small-$x$ extrapolation could affect the value of
the spin content, we compare our antiquark distribution with
some of the recent analyses. In Fig.\ref{fig:dqbar}, 
our AAC curve is shown together with the LSS (Leader-Sidorov-Stamenov)
and SMC (Spin Muon Collaboration) distributions.
The antiquark distribution is not explicitly calculated in the SMC paper,
so that the curve is obtained by transforming their distributions.
The SMC curve deviates from others in the large-$x$ region. 
However, the large-$x$ difference does not matter at this stage
because the antiquark distribution does not contribute to $g_1$ at large $x$.
It should be clarified for example by the Drell-Yan measurements at RHIC
in any case. Another important point is in the small-$x$ region.
Our distribution falls off rather slowly in comparison with others,
which results in the small quark spin content. Unfortunately,
available data are taken in the region $x>0.004$, and $x$ is
not small enough to determine $\Delta\bar q$ uniquely. It is also
still too far away from meaningful flavor decomposition
for $\Delta \bar q$.\cite{km} 

Due to the lack of small-$x$ data, we had better fix the small-$x$
behavior of $\Delta \bar q$. As theoretical guidelines, the Regge theory
and perturbative QCD could be used. First, the Regge model predicts
$g_1 (x)  \sim  x^{-\alpha}$ as $x\rightarrow 0$ with
the intercepts $\alpha$ of $a_1 (1260)$, $f_1 (1285)$,
and $f_1 (1420)$ trajectories. However, the intercepts are not
well known and it is usually assumed as $\alpha_{a_1}=-0.5 \sim 0$.
The small-$x$ functional form of the GRV98 is $x \, \bar q \sim x^{-0.14}$
at $Q^2=1$ GeV$^2$ according to our numerical estimate, so that
the Regge prediction is $\Delta\bar q/\bar q \rightarrow x^{1.1 \sim 1.6}$.
Second, the perturbative QCD could also suggest the small-$x$ behavior.
However, we have to assume that singlet and gluon distributions are
constants as $x\rightarrow 0$ at certain $Q^2 (\equiv Q_1^2)$.
Although it is not obvious whether such $Q_1^2$ exists, let us assume 
$Q_1^2=0.3\sim 0.5$ GeV$^2$. Then, the singlet distribution becomes
$x^{-0.12 \sim -0.09}$ as $x\rightarrow 0$, hence
$\Delta\bar q/\bar q \rightarrow x^{1.0}$.
The Regge and ``pQCD'' distributions fall off much faster than ours
at small $x$.

%%%%%%%%%%%%%%%%%%%%%%%%%%%%%%%% figure %%%%%%%%%%%%%%%%%%%%%%%%%%%%%%%%%%%%%%
\begin{wrapfigure}{r}{0.46\textwidth}
   \vspace{-0.3cm}
   \begin{center}
      \epsfig{file=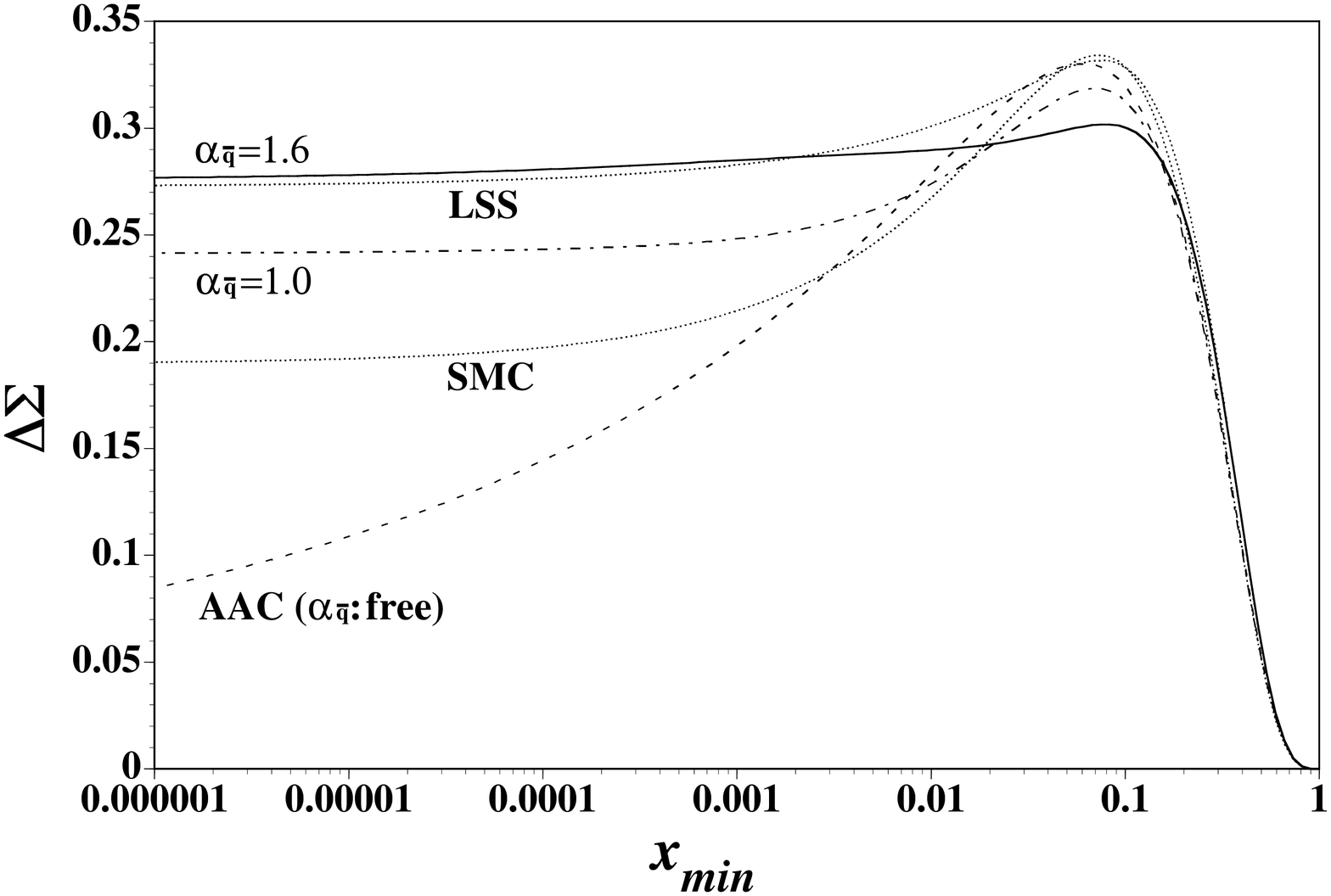,width=5.5cm}
   \end{center}
 \vspace{-0.75cm}
   \caption{\footnotesize Spin content
                          $\int_{x_{min}}^1 \Delta\Sigma dx$.}
\label{fig:dsigma}
\end{wrapfigure}
%%%%%%%%%%%%%%%%%%%%%%%%%%%%%%%% figure %%%%%%%%%%%%%%%%%%%%%%%%%%%%%%%%%%%%%%
From these theoretical suggestions, we analyzed the data again
by fixing the parameter $\alpha_{\bar q}$
($\Delta\bar q/\bar q \rightarrow x^{\alpha_{\bar q}}$).
First, we fixed the parameter at 1.0 which is suggested by
pQCD and it is also about the lower bound of the Regge model.
Second, it is taken as 1.6 which is the upper bound of the Regge.
Then, we obtain the NLO results:
\begin{alignat}{3}
\chi^2 = & 305.8, \ \ \ & \Delta\Sigma = & 0.241, \ \ \ 
                        & {\rm for}\ \alpha_{\bar q} = & 1.0 , 
\nonumber \\
\        & 323.5, \ \ \ & \              & 0.276, \ \ \ 
                        & \                            & 1.6 .   
\end{alignat}
The $\chi^2$ change for $\alpha_{\bar q}$=1.0
from the previous NLO value (300.4) is merely 1.8\%,
but we notice the large change in the spin content (5\%$\rightarrow$24\%).
Because of the small $\chi^2$ change, the $\alpha_{\bar q}$=1.0 fit
could be equally taken as a good solution.
On the other hand, the $\chi^2$ change is rather large for 
$\alpha_{\bar q}$=1.6. According to these results, the spin content is
24\%$\sim$28\% which is in the range of the widely-quoted values.
In this way, we find that the spin content is very sensitive to
the small-$x$ behavior of the polarized antiquark distribution
and that it cannot be fixed by the present $g_1$ data.

%%%%%%%%%%%%%%%%%%%%%%%%%%%%%%%%%%%%%%%%%%%%%%%%%%%%%%%%%%%%%%%%%%%%%%%%%%%%%%%
\section{Conclusions}\label{concl}

It is clarified in our $\chi^2$ analysis of the $A_1$ data
that the polarized antiquark distribution is not well determined
in the small- and large-$x$ regions. This fact leads to the conclusion that
the quark spin content is not well constrained only by the present
$A_1$ data. We need small- and large-$x$ measurements in future.
From our studies, we have proposed three sets of distributions:
LO, NLO-1 with free $\alpha_{\bar q}$, and NLO-2 with fixed
$\alpha_{\bar q}=1.0$.\cite{aac}

%%%%%%%%%%%%%%%%%%%%%%%%%%%%%%%%%%%%%%%%%%%%%%%%%%%%%%%%%%%%%%%%%%%%%%%%%%%%%%%
\section*{Acknowledgments}
S.K. was partly supported by the Grant-in-Aid for Scientific Research
from the Japanese Ministry of Education, Science, and Culture under
the contract number 10640277.
He would like to thank the theory division of CERN, where this manuscript
is partially written, for supporting his stay.

%%%%%%%%%%%%%%%%%%%%%%%%%%%%%%%%%%%%%%%%%%%%%%%%%%%%%%%%%%%%%%%%%%%%%%%%%%%%%%%

%%%%%%%%%%%%%%%%%%%%%%%%%%%%%%%%%%%%%%%%%%%%%%%%%%%%%%%%%%%%%%%%%%%%%%%%%%%%%%%


\section*{References}
\begin{thebibliography}{99}
\bibitem{aac}  Asymmetry Analysis Collaboration, Y. Goto {\it et al.},
                hep-ph/0001046 (Phys. Rev. D in press) and references 
                therein.
\bibitem{skpr} S. Kumano, Phys. Rep. {\bf 303}, 183 (1998).
\bibitem{q2-evol}   M. Miyama and S. Kumano,
                        Comput. Phys. Commun. {\bf 94}, 185 (1996);
                    M. Hirai, S. Kumano, and M. Miyama,
                        {\it loc. cit.} {\bf 108}, 38 (1998);
                                        {\bf 111}, 150 (1998).
\bibitem{km} S. Kumano and M. Miyama, Phys. Lett. B {\bf 497}, 149 (2000).
\end{thebibliography}
\end{document}